\documentclass[twocolumn,aps,prl] {revtex4} 
\usepackage{graphicx}
\begin{document}
\pagestyle{empty} 
\title{Leak-rate of seals: comparison of theory with experiment}
\author{  
B. Lorenz and B.N.J. Persson} 
\affiliation{IFF, FZ J\"ulich, D-52425 J\"ulich, Germany}

\begin{abstract}
Seals are extremely useful devices to prevent fluid leakage. 
We present experimental results for the leak-rate of rubber seals, and compare
the results to a novel theory, which is 
based on percolation theory and a recently developed contact mechanics theory.
We find good agreement between theory and experiment.
\end{abstract}
\maketitle


A seal is a device for closing a gap or making a joint fluid tight\cite{Flitney}.
Seals play a crucial role in many modern engineering devices, and the failure of
seals may result in catastrophic events, such as the Challenger disaster. 
In spite of its apparent 
simplicity, it is not easy to predict the leak-rate and
(for dynamic seals) the friction forces\cite{Mofidi} for seals. 
The main problem is the influence of surface
roughness on the contact mechanics at the seal-substrate interface. 
Most surfaces of engineering interest have surface roughness
on a wide range of length scales\cite{P3}, e.g, from cm to nm, which will influence the leak rate
and friction of seals, and accounting for the whole range of surface roughness
is impossible using standard numerical methods, such as the Finite Element Method.

\begin{figure}
\includegraphics[width=0.45\textwidth,angle=0]{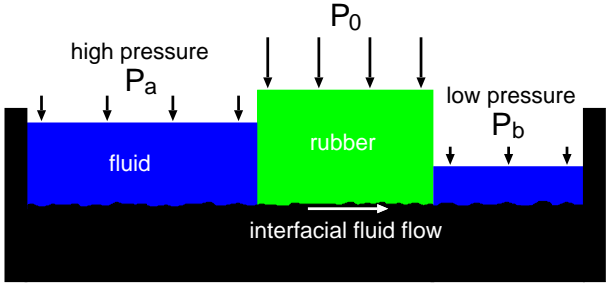}
\caption{\label{water.rubber}
Rubber seal (schematic). The liquid on the left-hand-side is under the hydrostatic
pressure $P_{\rm a}$ and the liquid to the right under the pressure $P_{\rm b}$
(usually, $P_{\rm b}$ is the atmospheric pressure).
The pressure difference $\Delta P = P_{\rm a}-P_{\rm b}$ results in liquid flow
at the interface between the rubber seal and the rough substrate surface. The
volume of liquid flow per unit time is denoted by $\dot Q$, and depends on the 
squeezing pressure $P_0$ acting on the rubber seal. 
}
\end{figure}

In this paper we present experimental results for the leak-rate of rubber seals,
and compare the results to a novel theory\cite{Creton,P3,Yang}, 
which is
based on percolation theory and a recently developed contact mechanics 
theory\cite{JCPpers,PerssonPRL,PSSR,P1,Bucher,YangPersson,PerssonJPCM}, 
which accurately takes into account the elastic 
coupling between the contact regions in the nominal rubber-substrate contact area.
Earlier contact mechanics models, such as the Greenwood--Williamson\cite{GW} model or the 
model of Bush et al\cite{Bush},
neglect this elastic coupling, which results in highly 
incorrect results\cite{Carlos,Carbone}, 
in particular for the relations between the squeezing pressure and the
interfacial separation\cite{Lorenz}. 
We assume that purely elastic deformation occurs in the solids, which is
the case for rubber seals.  

\begin{figure}
\includegraphics[width=0.45\textwidth,angle=0.0]{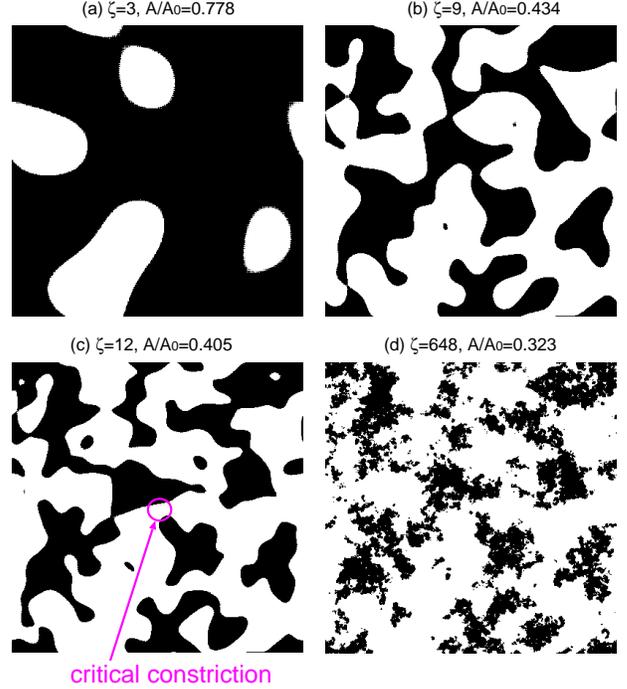}
\caption{\label{percolationpic}
The contact region at different magnifications $\zeta = 3$, 9, 12 and 648, are shown in
(a)-(d) respectively. 
When the magnification increases from 9 to 12 the non-contact region percolate.
At the lowest magnification $\zeta = 1$: $A(1)=A_0$. The figure is the result
of Molecular Dynamics simulations of the contact between elastic solids with randomly rough surfaces,
see Ref. \cite{Yang}.
}
\end{figure}

\vskip 0.3cm

Consider the fluid leakage through a rubber seal, from a high fluid pressure $P_{\rm a}$ region, to a
low fluid pressure $P_{\rm b}$ region, as in Fig. \ref{water.rubber}. 
Assume that the nominal contact region between the rubber and the hard countersurface is
rectangular with area $L_x\times L_y$. 
We assume that the high pressure fluid region is for $x<0$
and the low pressure region for $x>L_x$. We  ``divide'' the contact region into squares with
the side $L_x=L$ and the area $A_0=L^2$ (this assumes that $N=L_y/L_x$ is an integer, but this 
restriction does not affect the final result). 
Now, let us study the contact between the two solids within one of the squares
as we change the magnification $\zeta$. We define $\zeta= L/\lambda$, where $\lambda$ is the resolution.
We study how the apparent contact area (projected on the $xy$-plane),
$A(\zeta)$, between the two solids depends on the magnification $\zeta$.
At the lowest magnification we cannot observe any surface roughness, and 
the contact between the solids appears to be complete i.e., $A(1)=A_0$. 
As we increase the magnification
we will observe some interfacial roughness, and the (apparent) contact area will decrease.
At high enough magnification, say $\zeta = \zeta_{\rm c}$, a percolating path of 
non-contact area will be observed 
for the first time, see Fig.~\ref{percolationpic}. 
We denote the most narrow (and least high) constriction along this percolation path as
the {\it critical constriction}. The critical constriction will have the lateral
size $\lambda_{\rm c} = L/\zeta_{\rm c}$ and the surface separation at this point is denoted by 
$u_{\rm c}$. We can calculate $u_{\rm c} \approx u_1(\zeta_{\rm c})$,
using a recently developed contact mechanics theory\cite{YangPersson}. 
Thus, we define $u_1(\zeta)$ to be the (average) height separating the surfaces which appear to come into 
contact when the magnification decreases from $\zeta$ to $\zeta-\Delta \zeta$, where $\Delta \zeta$
is a small (infinitesimal) change in the magnification. $u_1(\zeta)$ can be calculated as described below.
As we continue to
increase the magnification we will find more percolating channels between the surfaces, but these will
have more narrow constrictions 
than the first channel which appears at $\zeta=\zeta_{\rm c}$, and as a first approximation we will
neglect the contribution to the leak-rate from these channels\cite{Yang}. 

A first rough estimate of the leak-rate is obtained by assuming that all the leakage 
occurs through the critical percolation channel, and that
the whole pressure drop $\Delta P = P_{\rm a}-P_{\rm b}$ (where $P_{\rm a}$ and $P_{\rm b}$ is the 
pressure to the left and right of the
seal) occurs over the critical constriction [of width and length $\lambda_{\rm c} \approx L/\zeta_{\rm c}$
and height $u_{\rm c}={u}_1 (\zeta_{\rm c})$]. If we approximate the critical constriction
as a pore with rectangular cross section (width and length $\lambda_c$ and height $u_c << \lambda_c$),  
and if assume an incompressible
Newtonian fluid, the volume-flow per unit time through the critical constriction
will be given by (Poiseuille flow) 
$$\dot Q = \alpha {u_1^3(\zeta_{\rm c}) \over 12 \eta}  \Delta P,\eqno(1)$$
where $\eta $ is the fluid viscosity. 
In deriving (1) we have assumed laminar flow and that $u_c << \lambda_c$,
which is always satisfied in practice. We have also assumed no-slip boundary condition
on the solid walls. This assumption is not always satisfied at the micro or nano-scale, but is likely to be
a very good approximation in the present case owing to surface roughness which occurs at length-scales 
shorter than the size of the critical constriction. 

In (1) we have introduced a factor $\alpha$ which depends on the
exact shape of the critical constriction, but which is expected to be of order unity. 
The flow rate expected for a channel with rectangular cross section (height $u_1$
and width and length $\lambda_{\rm c}$ with $u_1 << \lambda_{\rm c}$) 
correspond to $\alpha = 1$. However, the actual flow channel will not have a rectangular cross section
but the pore height must go continuously to zero at the ``edges'' in the direction perpendicular to the fluid flow.
In addition, the channel is of course not exactly rectangular in the $xy$-plane, and this too will effect $\alpha$.
Note also that a given percolation channel could have several narrow (critical or nearly critical)
constrictions of nearly the same dimension
which would reduce the flow along the channel. But in this case one would also expect more channels from
the high to the low fluid pressure side of the junction, which would tend to increase the leak rate.
These two effects will, at least in the simplest picture, compensate each other (see discussion in Ref. [5]).
Finally, since there are
$N=L_y/L_x$ square areas in the rubber-countersurface (apparent) contact area, we get the total leak-rate
$$\dot Q = \alpha {L_y \over L_x} {u_1^3(\zeta_{\rm c}) \over 12 \eta}  \Delta P.\eqno(2)$$ 

To complete the theory we must calculate the separation $u_{\rm c}=u_1(\zeta_{\rm c})$ 
of the surfaces at the
critical constriction. We first determine the critical magnification $\zeta_{\rm c}$ by assuming that the 
apparent relative contact area at this point is given by site percolation theory. 
Thus, the relative contact area $A(\zeta)/A_0 \approx 1-p_{\rm c}$, where $p_{\rm c}$  is the 
so called site percolation threshold\cite{Stauffer}. 
For an infinite-sized systems 
$p_{\rm c}\approx 0.696$ for a hexagonal lattice and $0.593$ for a square lattice\cite{Stauffer}. 
For finite sized systems the percolation will, on the average, occur for (slightly) smaller values
of $p$, and fluctuations in the percolation threshold will occur between 
different realizations of the same physical system. 
We take $p_{\rm c}\approx 0.6$ so that $A(\zeta_{\rm c})/A_0 \approx 0.4$ will determine the critical
magnification $\zeta=\zeta_{\rm c}$. 

The (apparent) relative contact area $A(\zeta)/A_0$ at the magnification $\zeta$
can be obtained using the contact mechanics 
formalism developed elsewhere\cite{PSSR,YangPersson,P1,Bucher,JCPpers},
where the system is studied at different magnifications $\zeta$.
We have\cite{JCPpers,PerssonPRL}

$${A(\zeta)\over A_0} = {1\over (\pi G )^{1/2}}\int_0^{P_0} d\sigma \ {\rm e}^{-\sigma^2/4G} 
= {\rm erf} \left ( P_0 \over 2 G^{1/2} \right )$$
where
$$G(\zeta) = {\pi \over 4}\left ({E\over 1-\nu^2}\right )^2 \int_{q_0}^{\zeta q_0} dq q^3 C(q)$$
where the surface roughness power spectrum
$$C(q) = {1\over (2\pi)^2} \int d^2x \langle h({\bf x})h({\bf 0})\rangle {\rm e}^{-i{\bf q}\cdot {\bf x}}$$
where $\langle ... \rangle$ stands for ensemble average. 
Here $E$ and $\nu$ are the Young's elastic modulus and the Poisson 
ratio of the rubber. The height profile $h({\bf x})$ of the rough surface can be measured routinely
today on all relevant length scales using optical and stylus experiments.

\begin{figure}
\includegraphics[width=0.35\textwidth,angle=0.0]{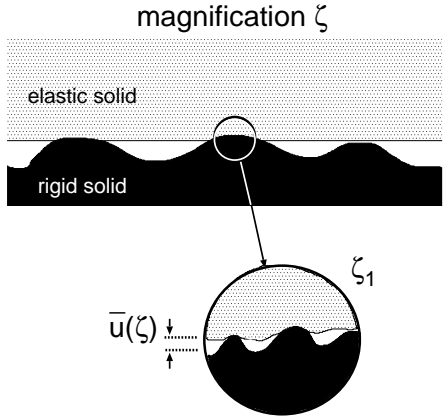}
\caption{\label{asperity.mag}
An asperity contact region observed at the magnification $\zeta$. It appears that
complete contact occur in the asperity contact region, but when the magnification is
increasing to the highest (atomic scale) magnification $\zeta_1$, 
it is observed that the solids are actually separated by the average distance $\bar{u}(\zeta)$.
}
\end{figure}

We define
$u_1(\zeta)$ to be the (average) height separating the surfaces which appear to come into 
contact when the magnification decreases from $\zeta$ to $\zeta-\Delta \zeta$, where $\Delta \zeta$
is a small (infinitesimal) change in the magnification. $u_1(\zeta)$ is a monotonically decreasing
function of $\zeta$, and can be calculated from the average interfacial separation
$\bar u(\zeta)$ and $A(\zeta)$ using
(see Ref.~\cite{YangPersson})
$$u_1(\zeta)=\bar u(\zeta)+\bar u'(\zeta) A(\zeta)/A'(\zeta).$$
The quantity $\bar u(\zeta)$ is the average separation between the surfaces in the apparent contact regions
observed at the magnification $\zeta$, see Fig.~\ref{asperity.mag}. 
It can be calculated from\cite{YangPersson}
$$\bar{u}(\zeta ) = \surd \pi \int_{\zeta q_0}^{q_1} dq \ q^2C(q) 
w(q)
 \int_{p(\zeta)}^\infty dp' 
 \ {1 \over p'} e^{-[w(q,\zeta) p'/E^*]^2},$$
where $p(\zeta)=P_0A_0/A(\zeta)$
and
$$w(q,\zeta)=\left (\pi \int_{\zeta q_0}^q dq' \ q'^3 C(q') \right )^{-1/2}.$$
The function $P(q,p,\zeta)$ is given by
$$P(q,p,\zeta) = {2\over \surd \pi} \int_0^{s(q,\zeta)p} dx \ e^{-x^2},$$
where $s(q,\zeta)=w(q,\zeta)/E^*$.

\begin{figure}
\includegraphics[width=0.22\textwidth,angle=0.0]{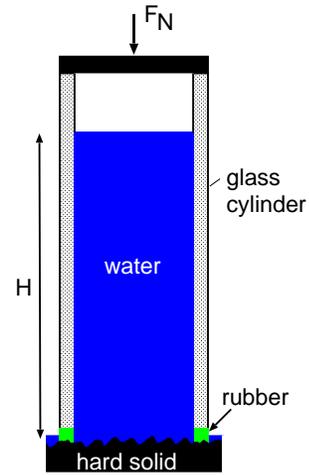}
\caption{\label{tube.water}
Experimental set-up for measuring the leak-rate of seals.
A glass (or PMMA) cylinder with a rubber ring attached to one end is squeezed against
a hard substrate with well-defined surface roughness. The cylinder is filled with 
water, and the leak-rate of the water at the rubber-countersurface
is detected by the change in the height of the water in the cylinder. 
}
\end{figure}

\begin{figure}
\includegraphics[width=0.45\textwidth,angle=0.0]{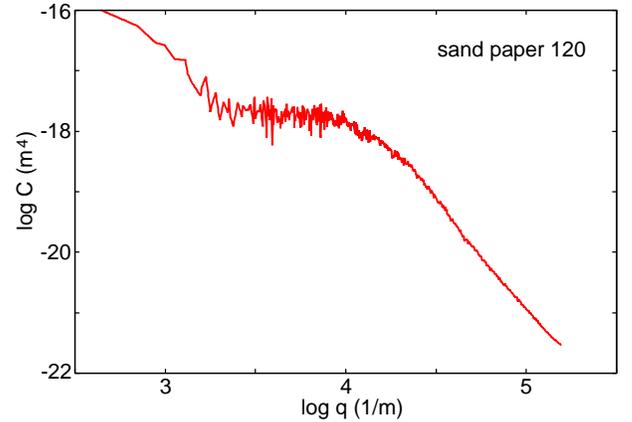}
\caption{\label{logq.logCq.sandPaper120}
Surface roughness power spectrum of sand paper 120. The surface has the root-mean-square roughness
$44 \ {\rm \mu m}$ and the surface area (including only the surface roughness with wavelength
above $\lambda_1 = 20 \ {\rm \mu m}$) is about $40 \%$ larger than the nominal 
surface area $A_0$ 
(i.e., the surface area projected on the $xy$-plane).
}
\end{figure}

We have performed a very simple 
experiment to test the theory presented above. 
In Fig.~\ref{tube.water} we show our 
set-up for measuring the leak-rate of seals.
A glass (or PMMA) cylinder with a rubber ring (with rectangular cross-section)
attached to one end is squeezed against
a hard substrate with well-defined surface roughness. The cylinder is filled with 
water, and the leak-rate of the fluid at the rubber-countersurface
is detected by the change in the height of the fluid in the cylinder. In this case
the pressure difference $\Delta P = P_{\rm a}-P_{\rm b} = \rho g H$, where $g$ is the gravitation
constant, $\rho$ the fluid density and $H$ the height of the fluid column. With $H\approx 1 \ {\rm m}$
we get typically $\Delta P \approx 0.01 \ {\rm MPa}$. With the diameter of the glass cylinder of
order a few cm, the condition $P_0>> \Delta P$ (which is necessary in order to be able to
neglect the influence on the contact mechanics from the fluid pressure at the rubber-countersurface)
is satisfied already for loads (at the upper
surface of the cylinder) of order kg. In our study we use a   
rubber ring with the Young's elastic modulus $E=2.3 \ {\rm MPa}$, and with the inner and outer diameter
$3 \ {\rm cm}$ and $4 \ {\rm cm}$, respectively, and the height $0.5 \ {\rm cm}$. 
The rubber ring was made from a silicon elastomer (PDMS) prepared using
a two-component kit (Sylgard 184) purchased from Dow Corning (Midland, MI). The kit consist of a base 
(vinyl-terminated polydimethylsiloxane) and a curing agent (methylhydrosiloxane-dimethylsiloxane copolymer) 
with a suitable catalyst. From these two components we prepared a mixture 10:1 (base/cross linker) in weight.
The mixture was degassed to remove the trapped air induced by stirring from the mixing process and then poured into casts. The bottom of these casts was made from glass to obtain smooth surfaces. The samples were cured in an oven at $80 ^\circ {\rm C}$ for 12 h.
The substrate is a corundum paper (grit size 120) with the root-mean-square roughness
$44 \ {\rm \mu m}$. 
From the measured surface topography we obtain the surface roughness power spectrum $C(q)$
shown in Fig. \ref{logq.logCq.sandPaper120}.  

\begin{figure}
\includegraphics[width=0.45\textwidth,angle=0.0]{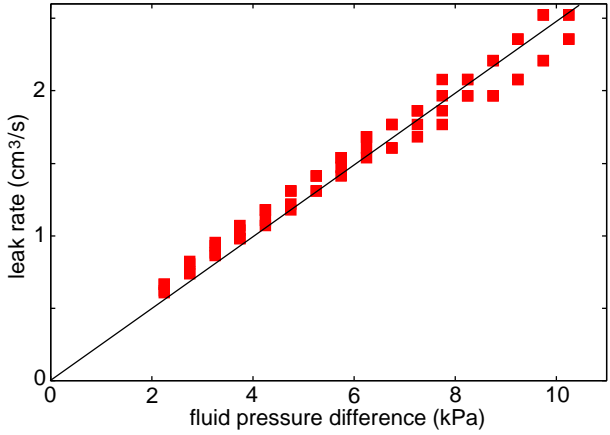}
\caption{\label{pressurelealrate}
Square symbols: the measured leak rate for different fluid pressure drop $\Delta P = P_{\rm a}-P_{\rm b}$ 
for the nominal squeezing pressure $P_0 \approx 60 \ {\rm kPa}$. 
}
\end{figure}

\begin{figure}
\includegraphics[width=0.45\textwidth,angle=0.0]{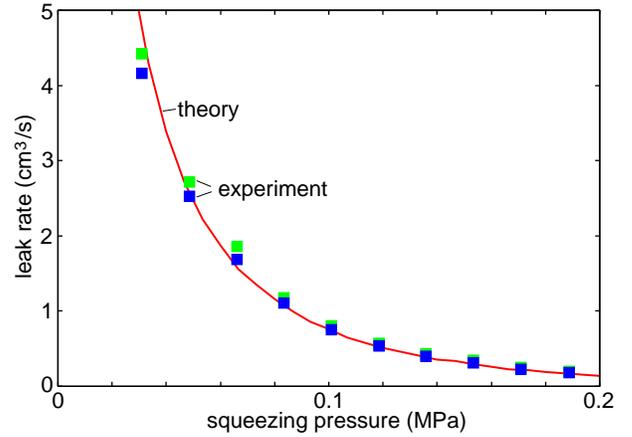}
\caption{\label{pressure.leakrate}
Square symbols: the measured leak rate for ten different squeezing pressures. 
The experiment was performed twice, corresponding to the two data points for each pressure.
Solid line:
the calculated leak rate using the measured surface topography, the measured rubber elastic
modulus $E=2.3 \ {\rm MPa}$ and the fluid pressure difference 
$\Delta P = P_{\rm a}-P_{\rm b} = 10 \ {\rm kPa}$ obtained from the height of the water column.  
We have used $\alpha = 0.2$.
}
\end{figure}

\begin{figure}
\includegraphics[width=0.45\textwidth,angle=0.0]{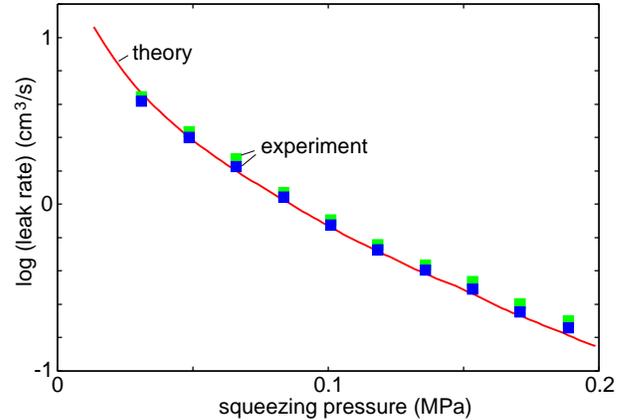}
\caption{\label{pressure.logLeakRate}
The same as in Fig. \ref{pressure.leakrate} but for a logarithmic (with 10 as basis)
leak-rate scale.
}
\end{figure}

\begin{figure}
\includegraphics[width=0.45\textwidth,angle=0.0]{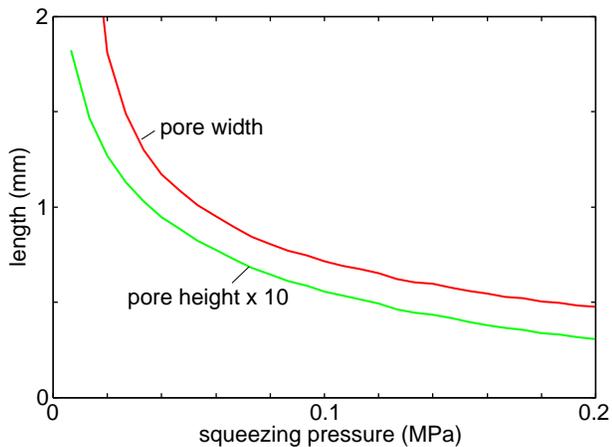}
\caption{\label{pressure.poresize}
Calculated critical pore size as a function of the squeezing pressure.
For the same system as in Fig. \ref{pressure.leakrate}.
}
\end{figure}

According to (1) we expect the leak-rate to increase linearly with the fluid pressure difference
$\Delta P = P_{\rm a}-P_{\rm b}$. We first performed some experiments to test this prediction. In Fig. \ref{pressurelealrate}
we show the measured leak rate for different fluid pressure drop $\Delta P$ 
for the nominal squeezing pressure $P_0 \approx 60 \ {\rm kPa}$. To within the accuracy of the experiment, the leak-rate depends linearly on
$\Delta P$.  

In Fig. \ref{pressure.leakrate} we show the measured leak rate for ten different squeezing pressures
(square symbols).
The solid line is the calculated leak rate using the measured rubber elastic modulus $E=2.3 \ {\rm MPa}$ and the 
surface power spectrum $C(q)$ shown in Fig. \ref{logq.logCq.sandPaper120}.
In Fig. \ref{pressure.logLeakRate} we show 
the same as in Fig. \ref{pressure.leakrate} but for a logarithmic (with 10 as basis)
leak-rate scale.

\begin{figure}
\includegraphics[width=0.45\textwidth,angle=0.0]{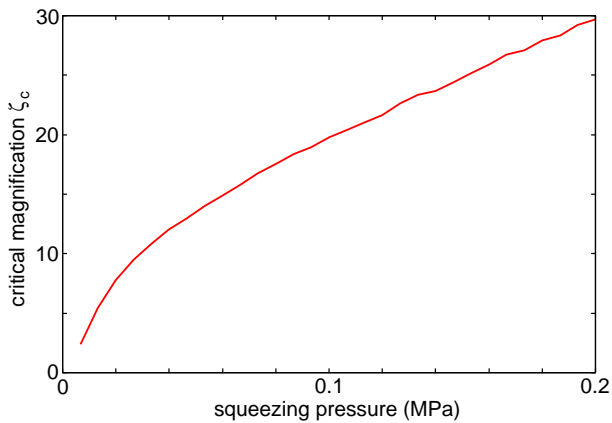}
\caption{\label{pressure.magnification}
Calculated critical magnification $\zeta_c$, where the non-contact area percolate, 
as a function of the squeezing pressure.
For the same system as in Fig. \ref{pressure.leakrate}. 
}
\end{figure}

In Fig. \ref{pressure.poresize} we show the 
calculated critical pore size as a function of the squeezing pressure. Note that the height of the critical
pore is about 10 times smaller than the lateral size of the pore. Finally, in Fig. \ref{pressure.magnification}
we show the critical magnification $\zeta_c$, where the non-contact area percolate,
as a function of the squeezing pressure. Note that, as expected, 
the percolation of the non-contact area occur at
higher and higher magnification as the squeezing pressure increases.

Sand paper has much sharper and larger roughness than the counter surface used in normal rubber seal applications.
However, from a theory point of view it should not really matter on which length scale the
roughness occurs, except for ``complications'' such as the influence of adhesion and fluid
contamination particles (which tend to clog the flow channels). 
Nevertheless, the theory assumes that the average surface slope is not too large and
we plan to measure the leak rate for rubber seal in contact with sand blasted Plexiglas
with a root-mean-square roughness in the micrometer range.
Our initial experiment
with Plexiglas showed that the leak rate decreased by time and finally no leaking could be observed.
But this experiment used unfiltered tap water which contains contamination particles which clogged the channels.
We are now using distilled water and find the leak rate (for a given fluid pressure difference) to be practically time
independent. But these are still preliminary studies and we will report on the final results elsewhere.

Finally, we note that it is nearly impossible to calculate $\alpha$ 
theoretically and we have fitted $\alpha$ to reproduce the experimental data.
However, this is just
one parameter and the dependence of the leak rate on the nominal rubber-countersurface pressure is highly non-trivial and
accurately given by the theory. Also, $\alpha$ is of order unity as expected from theory. The only way (as we see it)
to obtain a theory-estimate of $\alpha$ would be to generate critical constrictions using numerically exact contact
mechanics calculations (e.g., molecular dynamics), and to simulate the fluid flow through the constrictions using numerical methods of fluid
flow dynamics.  

To summarize, we have compared 
experimental data with theory for the leak-rate of seals. The theory is 
based on percolation theory and a recently developed contact mechanics theory.
The experiments are for silicon rubber seals in contact with sand paper. The
elastic properties of the rubber and the surface topography of the 
sand paper are fully characterized.
The calculated leak-rate $\dot Q$ is in good agreement with experiment.
The theory only account for fluid flow through the percolation channels
observed at (or close to) the percolation threshold. A more accurate treatment should include
also flow channels observed at higher magnification. This problem has similarities to 
current flow in random resistor networks\cite{Stauffer,Wu}.

\vskip 0.3cm

We thank Christian Schulze (ISAC, RWTH Aachen University) 
for help with the measurement of the surface topography of the sand
paper surfaces.
This work, as part of the European Science Foundation EUROCORES Program FANAS, was supported from funds 
by the DFG and the EC Sixth Framework Program, under contract N ERAS-CT-2003-980409.

\end{document}